\documentclass[amsmath,amssymb,12pt,floatfix] {revtex4}
\usepackage[dvips]{graphicx}
\usepackage{bm}

\newcommand{\be}{\begin{equation}}
\newcommand{\ee}{\end{equation}}
\newcommand{\ber}{\begin{eqnarray}}
\newcommand{\eer}{\end{eqnarray}}

\begin{document}
\title{Chemotaxis When Bacteria Remember: Drift versus Diffusion}
\author{Sakuntala Chatterjee(1), Rava Azeredo da Silveira(2,3) and Yariv Kafri(1)}
\affiliation{(1) Department of Physics, Technion, Haifa-32000, Israel \\
(2) Department of Physics and Department of Cognitive Studies, \'Ecole Normale
Sup\'erieure, 24, rue Lhomond, 75005 Paris, France \\
(3) Laboratoire de Physique Statistique, Centre National de la Recherche Scientifique, Universit\'e Pierre et Marie Curie, Universit\'e Denis Diderot,
 France }
\begin{abstract}
{\it {\sl Escherichia coli} ({\sl E. coli}) bacteria govern their trajectories by 
switching between running and tumbling modes as a function of the nutrient 
concentration they experienced in the past. At short time one observes a drift
 of the bacterial population, while at long time one observes accumulation in 
high-nutrient regions. Recent work has  
 viewed chemotaxis as a compromise between drift 
toward favorable regions and accumulation in favorable regions. A number of 
 earlier studies assume that a bacterium resets its memory at tumbles -- a fact
 not borne out by experiment -- and make use of approximate coarse-grained
 descriptions. Here, we revisit the problem of chemotaxis without resorting to
 any memory resets. We find that when bacteria respond to the environment in
 a non-adaptive manner, chemotaxis is generally 
dominated by diffusion, whereas when
 bacteria respond in an adaptive manner, chemotaxis is dominated by a bias in
 the motion. In the adaptive case, favorable drift
 occurs together with favorable accumulation.
 We derive our results from detailed 
simulations and a variety of analytical arguments. In particular, we introduce
 a new coarse-grained description of chemotaxis as biased diffusion, and we
 discuss the way it departs from older coarse-grained descriptions. }
\end{abstract}
\maketitle

\section*{Author Summary}
The chemotaxis of {\sl Escherichia coli} is a prototypical model of navigational strategy. The bacterium maneuvers by switching between near-straight motion, termed runs, and tumbles which reorient its direction. To reach regions of high nutrient concentration, the run-durations are modulated according to the nutrient concentration experienced in recent past. This  navigational strategy is quite general, in that
the mathematical description of these modulations also accounts for the active motility of {\sl C. elegans}
and for thermotaxis in {\sl  Escherichia coli}. Recent studies have pointed to a possible incompatibility  between reaching regions of high nutrient concentration quickly and staying there at long times. We use  numerical investigations and analytical arguments to reexamine navigational strategy in bacteria. We show that, by accounting properly for the full memory of the bacterium, this paradox is resolved. Our work clarifies the mechanism  that underlies chemotaxis and indicates that chemotactic navigation in wild-type bacteria is controlled by drift while in some mutant bacteria it is controlled  by a modulation of the diffusion.  We also propose a new set of effective, large-scale equations which describe
bacterial chemotactic navigation. Our description is significantly different from previous ones, as it results from a conceptually different coarse-graining procedure.

\section*{Introduction}
The bacterium {\sl E. coli} moves by switching between two types of motions, 
termed `run' and `tumble' \cite{bergbook}. Each results from a distinct
movement of the
 flagella. During a run, flagella motors rotate counter-clockwise 
(when looking at the bacteria from the back), inducing an almost constant forward 
 velocity of about
 $20 \mu m / s$, along a near-straight line. In an environment with
uniform nutrient concentration,  
run durations are distributed exponentially with a mean value of about
 $\tau_R= 1 s$ \cite{berg72}. When motors turn clockwise, the bacterium undergoes
 a tumble, during which, to a good approximation, it 
 does not translate but instead changes its direction randomly. In a uniform 
nutrient-concentration profile, 
the tumble duration is also distributed exponentially but
 with a much shorter mean  value of about $\tau_T=0.1 s$ \cite{turner}.

\begin{figure}[!ht]
\includegraphics[scale=1.0,angle=0]{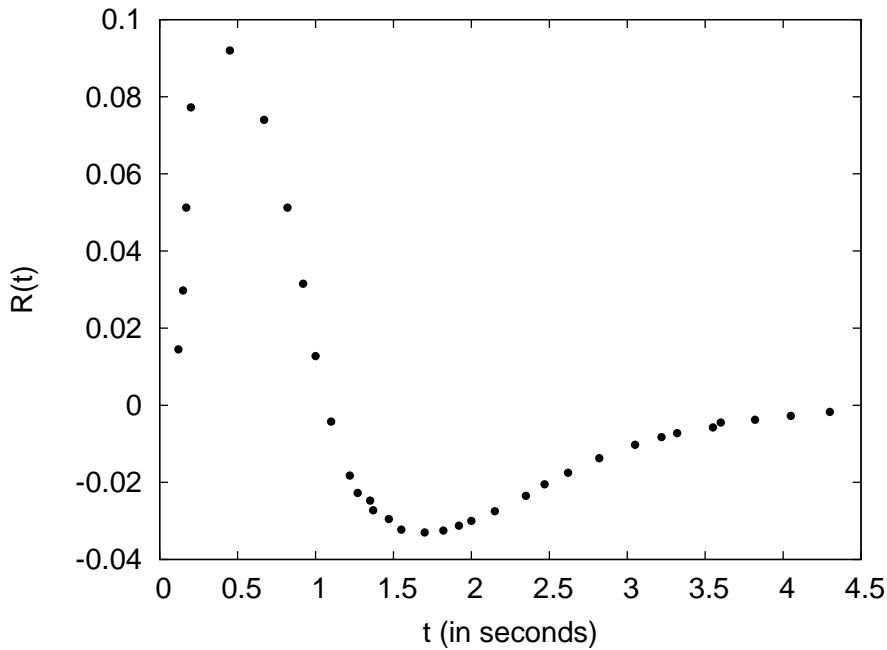}
\caption{ Bilobe response function of wild-type {\sl E. coli} used 
in the numerics in Fig. \ref{fig:bilin}. For the sake of 
 computational simplicity, we have used a 
discrete sampling of the experimental data presented in Ref. \cite{segall} instead
of working with the complete data set. This did not affect our conclusions.}
\label{fig:r}
\end{figure}

When the nutrient (or, more generally, chemoattractant)
concentration varies in space, bacteria tend to accumulate in
 regions of high concentration \cite{footnote,adler}. 
 This is achieved through a modulation of the run durations. 
The biochemical pathway that controls flagella dynamics is
well understood \cite{bergbook,eisen,leibler,uri} and the stochastic
`algorithm' which governs 
the behavior of a single motor is experimentally measured. The latter is routinely used
as a model for the motion of a bacteria with many motors 
\cite{bergbook,gennes,kafri,verg,lars}.
This algorithm represents the motion of the bacterium as a non-Markovian random walker whose stochastic run durations 
are modulated via a memory kernel, shown in Fig. \ref{fig:r}. Loosely speaking,
the kernel compares the nutrient concentration experienced in the recent 
past with that
experienced in the more distant past. If the difference 
is positive, the run duration is extended; if it is negative, the run duration is shortened.

In a complex medium bacterial navigation involves further
complications; for example,
interactions among the bacteria, and degradations or other dynamical
variations in the chemical environment. 
These often give rise to interesting collective
behavior such as pattern formation \cite{cates_pnas,
mittal}. However, in an  attempt to understand collective behavior, it is
imperative to first have at hand
 a clear picture of the behavior of a single bacterium in an
inhomogeneous chemical environment. We are concerned with this narrower
question in the present work.

Recent theoretical studies of single-bacterium
behavior have 
shown that a simple connection between the stochastic algorithm of motion and the
 average chemotactic response is far from obvious \cite{gennes,lars,kafri,verg}.
 In particular, it appeared 
that favorable chemotactic drift could not be
reconciled with  
 favorable accumulation at long times, and chemotaxis was viewed 
as resulting from
a compromise between the two \cite{lars}. The optimal nature of this
compromise in bacterial chemotaxis was examined in Ref. \cite{verg}. 
 In various approximations, while the
negative part of the response kernel was key to favorable
 accumulation in the steady state, it  suppressed
 the drift velocity. Conversely,
the positive part of the response kernel enhanced the drift velocity but
reduced the magnitude of the chemotactic response in the steady state.

Here, we carry out a detailed study of the 
chemotactic behavior of a single bacterium in one dimension.  We find that,
for an `adaptive' response kernel ({\sl i.e.}, when the positive
and negative parts of the response kernel have equal weight such that the
total area under the curve vanishes),
there is no incompatibility between a strong steady-state chemotaxis and a 
large drift
velocity. A strong steady-state chemotaxis occurs when the
positive peak of the response kernel occurs at a time much smaller than 
$ \tau_R$ and the negative peak at a time much larger than $\tau_R$, in
line with experimental observation. Moreover, we  obtain that the drift
velocity is also large in this case. For a general
`non-adaptive' response kernel ({\sl i.e.}, when the area under the response
kernel curve is non-vanishing), however, we find that a large drift velocity indeed
opposes chemotaxis. Our calculations show that, in this case,
 a position-dependent
diffusivity is responsible for chemotactic accumulation.

In order to explain our numerical results, we propose a simple 
coarse-grained model which describes the
bacterium as a biased random walker with a drift velocity and 
diffusivity, both of which are, in general,  position-dependent. This simple 
model yields good agreement with results of detailed  simulations. We 
emphasize that our model is distinct
from existing coarse-grained descriptions of  {\sl E. coli}
chemotaxis \cite{schnitzer,coarse,mittal,blythe}.  In these,  
coarse-graining was performed over  left- and right-moving bacteria 
separately,  after which the two resulting 
coarse-grained quantities  were then added to obtain an
equation for the total coarse-grained density. We point out why such
 approaches can fail and discuss the differences
between earlier models and the present coarse-grained model.

\section*{Models}
Following earlier studies of chemotaxis \cite{berg82,kafri}, we model the
navigational behavior of a bacterium by a stochastic law of motion with
Poissonian run durations. 
 A switch from run to tumble occurs during the
small time interval between $t$ and $t+dt$ with a probability 
\begin{equation}
\frac{dt}{\tau _{R}}\left\{ 1-\mathcal{F}\left[ c\right]\right\}.  
\label{eq:tumbfreq}
\end{equation}
Here, $\tau_R \simeq 1 s$ and 
$\mathcal{F}\left[ c\right] $ is a functional of the chemical concentration, 
$c(t^{\prime })$, experienced by the bacterium at times $t^{\prime }\leq t$.
In shallow nutrient gradients, the functional can be written as
 \begin{equation}
\mathcal{F}\left[ c\right] =\int_{-\infty }^{t}dt^{\prime
}R(t-t^{\prime })c(t^{\prime }) 
 \label{eq:linearresponse}
\end{equation}
 The response kernel, $R(t)$, encodes the action of the biochemical 
machinery that processes input signals from the
environment. Measurements of the change in the rotational bias of a flagellar
motor in wild-type bacteria, in response to instantaneous chemoattractant
pulses were reported in Refs. \cite{berg82,segall}; experiments were carried
out with a tethering assay. The response kernel obtained
from these measurements has a bimodal shape,
with  a positive peak around $t\simeq 0.5 s $ and a negative peak around 
$t\simeq 1.5 s $ (see Fig. \ref{fig:r}). The negative lobe is shallower
 than the positive one and extends up to $t\simeq 4 s$, beyond which
 it vanishes. The total area under the response curve is close to zero.
As in other studies of {\sl E. coli} chemotaxis, we take this response
kernel to describe the modulation of run duration of 
swimming bacteria \cite{gennes,lars,kafri,verg}. Recent experiments suggest
that tumble durations are not modulated by the chemical environment and that as
long as tumbles last long enough to allow for the reorientation of the cell, 
 bacteria can perform chemotaxis successfully \cite{bai,mora}.

  The model defined by Eqs. \ref{eq:tumbfreq} and
\ref{eq:linearresponse} is linear. Early experiments pointed to 
 a non-linear, in effect a threshold-linear, behavior of a  bacterium 
in response to chemotactic inputs \cite{berg82,segall}. In these studies, 
a bacterium modulated its motion in response to a positive
chemoattractant gradient, but not to a negative one. 
In the language of present model,
such a threshold-linear response entails replacing the functional defined in Eq.
\ref{eq:linearresponse} by zero whenever the integral is 
negative. More recent experiments suggest a different picture, in which
a non-linear response is expected only for a strong input
signal whereas the response to  weak chemoattractant gradient is well
described by a linear relation \cite{shimizu}. 
Here, we present an analysis of the linear model. For the sake of
 completeness, in Supporting Information, we 
present a discussion of models which include tumble modulations and a 
non-linear response kernel. Although recent experiments have ruled out the
existence of both these effects in {\sl E.coli} chemotaxis, in general 
such effects can be relevant to other systems with  similar forms of
 the response function.

The shape of the response function hints to a simple mechanism for
 the bacterium to reach regions with high nutrient concentration. 
The bilobe kernel measures a temporal
gradient of the nutrient concentration. According to Eq. \ref{eq:tumbfreq},
if the gradient is positive, runs are extended; if it is negative, runs
are unmodulated. However, recent literature \cite{gennes,lars,kafri} has
pointed out that the connection between this simple  picture and a detailed
quantitative analysis is tenuous. For example, de
 Gennes used Eqs. \ref{eq:tumbfreq} to calculate the chemotactic 
drift velocity of bacteria \cite{gennes}. He found that  
a singular kernel, $R(t)=\alpha \delta(t-\Delta)$, where $\delta$ is a Dirac
function and $\alpha$ a positive constant, 
lead to a mean velocity in the direction of increasing nutrient concentration 
even when bacteria are memoryless ($\Delta =0$). Moreover, any addition of a
 negative contribution to 
the response kernel, as seen in experiments (see Fig. \ref{fig:r}),
 lowered the drift velocity. Other studies considered
 the steady-state density profile of bacteria in a container with
closed walls, both in an
 approximation in which correlations between run durations and probability density 
were ignored \cite{lars} and in an approximation in which the memory of the 
bacterium was reset at run-to-tumble switches \cite{kafri}.
 Both these studies found that, in the steady state, a 
negative contribution to the response function was mandatory for bacteria
 to accumulate in regions of high nutrient concentration. These results
seem to imply that the  joint  requirement of
favorable transient drift and steady-state accumulation is 
problematic. The paradox was further complicated by the observation 
\cite{kafri}
that the steady-state single-bacterium probability density
 was sensitive to the precise shape of the kernel: when the negative part of 
the kernel
 was located far beyond $\tau_R$ it had little influence on the steady-state
 distribution \cite{lars}. In fact, for kernels  similar
 to the experimental one,  model
 bacteria accumulated in regions with low nutrient concentration in
 the steady state \cite{kafri}. 

\section*{Results}

\subsection*{Simulations and analytical treatment of chemotactic bacterial
accumulation}
In order to resolve these paradoxes and to better understand the mechanism
that leads to favorable accumulation of bacteria, 
we perform careful numerical studies of  bacterial
 motion in one dimension. In conformity with
experimental observations \cite{berg82,segall}, 
we do not make any assumption of memory reset at run-to-tumble switches.

We model a bacterium as a one-dimensional non-Markovian random
walker. The walker can move either to the left or to the
 right with a fixed speed,
$v$, or it can tumble at a given position before initiating a new run.
In the main paper, we present results only for the case of 
instantaneous tumbling with $\tau_T =0$, while results for non-vanishing
$\tau_T$ are discussed in Supporting Information. There, we
verify that for an adaptive response kernel $\tau_T$ does not have
any effect on the steady-state density profile. For a non-adaptive response
kernel, the  correction in the steady-state slope
  due to finite $\tau_T$ is small and 
 proportional to $\tau_T /\tau_R$.

The run durations are Poissonian and the tumble probability is given by 
Eq. \ref{eq:tumbfreq}.  The probability to change the run direction
after a tumble is assumed to have a fixed value, $q$, which we treat as a
parameter. The specific choice of the value of $q$ does not
affect our broad conclusions. We find that, as long as $q \neq 0$, only certain
detailed quantitative aspects of our numerical results depend on $q$. (See
Supporting Information for details on this point.) We assume that  
bacteria are in a box of size $L$ with reflecting walls and that they do not
interact among each other. We focus on the steady-state behavior of a
population. Reflecting boundary conditions are a 
simplification of the actual behavior \cite{angelani,galajda};
as long as the total `probability current' (see discussion below)
 in the steady state vanishes, our results remain valid even if the walls are
not reflecting.

As a way to probe chemotactic accumulation, 
we consider a linear concentration profile of nutrient: $c(x)=cx$.
 We work in a weak gradient limit, {\sl i.e.}, the value of
$\alpha c$ is chosen to be sufficiently small to allow for a 
 linear response. Throughout,
 we use $c=1/L$ in our numerics. From the linearity of the
problem, results for a different attractant gradient, $k/L$, can be obtained
from our results through a scaling factor $k$. In the linear reigme, we obtain
a spatially linear steady-state distribution of individual
bacterium positions, or, equivalently, a linear density profile of a
bacterial population. Its slope, which we denote by $\beta$, 
 is a measure of the strength of chemotaxis. A large slope indicates strong
bacterial preference for regions with higher nutrient
concentration. Conversely, a vanishing slope implies that bacteria
are insensitive to the gradient of nutrient concentration and are
equally likely to be anywhere along the line.
 We would like to understand the way in which the slope $\beta$
depends on the  different time scales present in the system.

\subsubsection*{Results with non-adaptive response kernels}

One particular advantage of a linear model is that a general
  problem can be solved by superposing the
solutions of simpler problems---namely, with delta-function response
kernels---with suitably chosen coefficients. Thus, solving the problem
with a singular response kernel amounts to a full solution and we focus here
on this case. 

In our simulations, we start from an arbitrary bacterium position inside
a  box of size $L$. Each time step has a duration $dt$, during which a
running bacterium
moves over a distance $vdt$. This distance corresponds to one lattice
spacing in our model, in which a lattice is introduced because time is
discretized. Throughout the numerics, we use $dt=0.01s$ and
$v=10 \mu m/s$, which means that the lattice spacing in our simulations is
 $0.1 \mu m$. Results for different 
values of $v$ can be obtained by rescaling the lattice spacing accordingly.
At the end of each time step, we compute the functional defined 
in Eq. \ref{eq:linearresponse}; for a singular response kernel, $R(t) = \alpha
\delta (t-\Delta)$, this 
takes the form $\alpha c\left [ x(t-\Delta)\right ]$, where $c\left [
x(t-\Delta)\right ]$ is the nutrient concentration experienced by the
bacterium at  time $t-\Delta$. At the end of each time step
the bacterium either tumbles, with a probability 
$\left ( 1- \alpha c\left [ x(t-\Delta)\right ] \right ) dt/\tau_R$,
or continues to move in the same direction. At every tumble, the velocity
of the bacterium is reversed with a probability $q$. 

 The system reaches a
steady state over a time scale which is of order $ L^2/D$, where the
diffusivity is given by $D = v^2 \tau_R$.
We verify numerically that after this
time the bacterial density profile inside the box does not change further and
assumes a time-independent linear form.
We focus on the slope, $\beta$, of this profile.
For an experimental realization of the steady-state behavior of
a single bacterium, we provide here an estimate of the time scales and length
scales involved.
Since the long-time behavior of the system is diffusive (see the discussion
of the coarse-grained model below), the relaxation time is  
$L^2/D$. Our results on the steady-state distribution of  bacteria hold,
realistically, if this relaxation time does not exceed
the typical  division time of an {\sl E. coli} bacterium,
 which is of the order of $30$ minutes. Substituting experimental values for
 the parameters, we find the description should be valid for system sizes $L
\lesssim 400 \mu m $. In our simulations, we use a somewhat larger system
($L=1000 \mu m$) so as to have cleaner results with negligible
 effects of the reflecting walls at the
two boundaries. (Numerics data show that the 
width of the boundary layer is about $\sim 80 \mu m$.)

\begin{figure}[!ht]
\includegraphics[scale=1.0,angle=0]{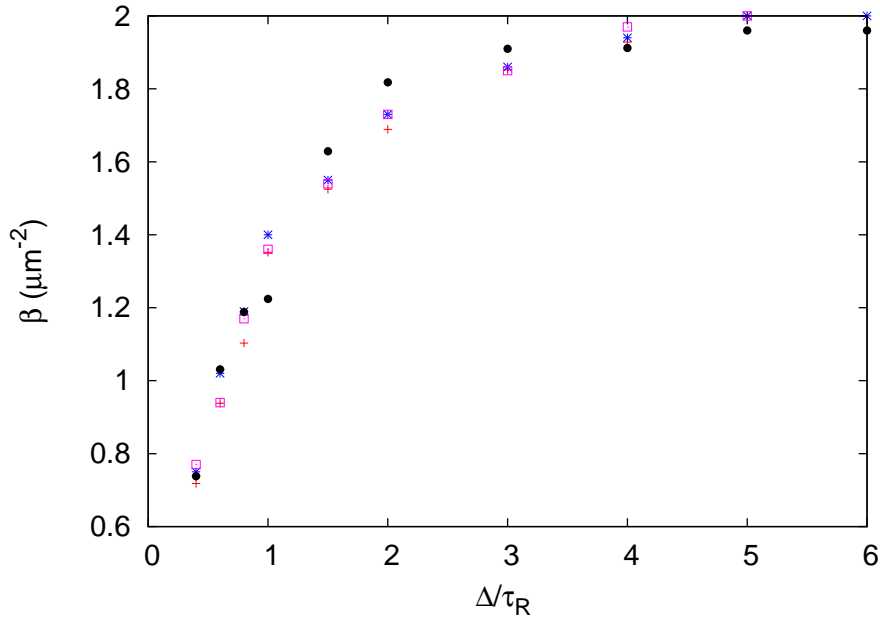}
\caption{ The slope $\beta$ (scaled by a factor of $10^8$) 
 as a function of   $\Delta/\tau_R $, for the choice of response kernel
$R(t)=\alpha \delta (t-\Delta)$. Note that for $\Delta \gg \tau_R$ the slope
saturates to a non-vanishing value. The symbols
$+$, $\ast$, and $\square$ correspond to slopes measured in simulations with 
$\tau_R=1.5, 1.0$, and $0.5$ seconds, respectively. The black solid circles are
derived from our coarse-grained formulation (Eq. \ref{eq:slope}).
Here $q=0.5$, $\alpha = -0.02$, $L=1000 \mu m$, $c=0.001 \mu m ^{-1}$,
 $v=10 \mu m /s$.}
\label{fig:satq5}  
\end{figure}

According to our numerical simulations, for $\alpha <0$, $\beta$ {\it increases}
 with $\Delta$ and displays a plateau for $\Delta \gg
\tau_R$ (Fig. \ref{fig:satq5}). Simulations probing various values of 
 $\tau_R$ also confirmed that  $\beta = F (\Delta / \tau_R)$, {\sl i.e.}, that
the slope is a scaling function of $\Delta /\tau_R$. Clearly, for positive $\alpha$
the sign of $\beta$ is simply reversed, which corresponds to
 an unfavorable chemotaxis \cite{schnitzer,lars}.  

For small $\Delta$, one can write down an approximate master equation for
left-mover and right-mover densities and use it to show that the slope increases
linearly with $\Delta$ (see Supporting Information for details).  
It is surprising, however, that the slope appears to saturate to a non-vanishing
value for $\Delta \gg
\tau_R$. Indeed one would expect that, if the response
kernel relies on a time much  earlier than $t-\tau_R$,  
a large enough number of tumbles occur between this past time and the present
time so as to eliminate
 any correlation between the nutrient concentration in the
past and the present direction of motion. 
If this argument holds, one would expect that the slope $\beta$
vanish for $\Delta \gg \tau_R$. Below, we return to this argument and
explain why it is misleading.

\subsubsection*{Results with adaptive response kernels}
For wild-type bacteria, the total area under the response kernel vanishes
 (Fig. \ref{fig:r}). As a result, their behavior is adaptive: chemotaxis is
insensitive to the overall level of nutrient, but sensitive to spatial
variations \cite{segall, berg82}. 
In this section, before examining the case of a
 bilobe response kernel 
similar to the experimental one, we consider a toy model defined by
the difference of two singular forms:
 $R(t)=\alpha \delta (t-\Delta_1) - \alpha \delta (t-\Delta_2)$, with $\alpha
>0$. Because our problem is linear, the steady-state slope of bacterial
density, $\beta$, can be calculated from a simple linear superposition, as: 
\be
\beta =  F \left ( \frac{\Delta_1}{\tau_R} \right ) - 
F \left ( \frac{\Delta_2}{\tau_R} \right ).
\label{eq:f2del}
\ee  
Since the function $F(\cdot)$ is monotonic, the absolute value of $\beta$ 
increases with the difference of $\Delta_1$ and $\Delta_2$. Strong chemotaxis
occurs when  $\Delta_1=0$ and $\Delta_2 \gg \tau_R$.

\begin{figure}[!ht]   
\includegraphics[scale=1.0,angle=0]{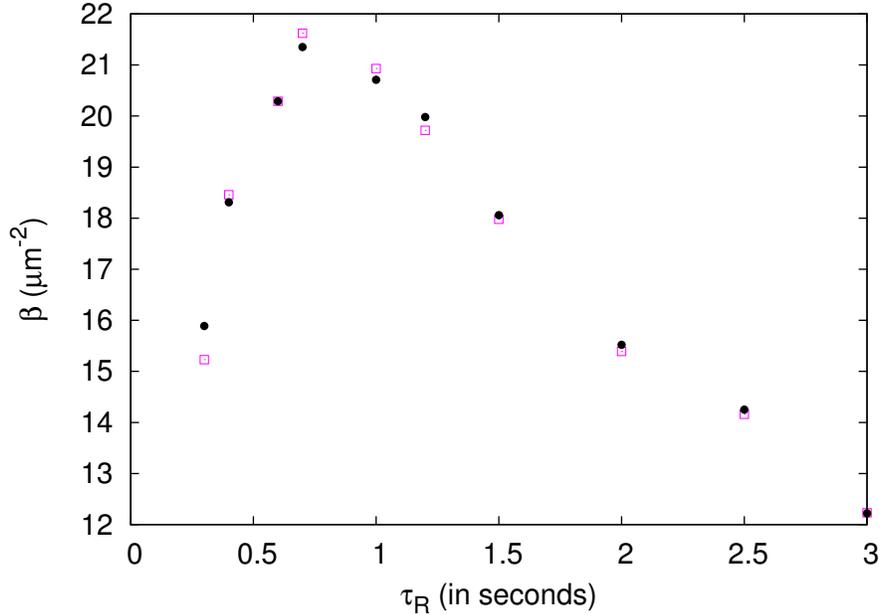}
\caption{ The slope $\beta$ (scaled by a factor of $10^8$)
 as a function of  $\tau_R$ for the experimental 
response kernel shown in Fig. \ref{fig:r}. Open squares: numerical results from
simulations. Solid circles: prediction of the coarse-grained model. Here,
 $q=0.4$, $L=1000 \mu m$, $c=0.001 \mu m ^{-1}$, $v=10 \mu m /s$.}
\label{fig:bilin}
\end{figure}

We now turn to the experimental case of a bilobe response kernel.  
It is not computationally 
feasible to work with the complete set of experimental data \cite{segall}, 
so we have used a discrete subset (Fig. \ref{fig:r}) which we represent
as 
a series of delta-functions. Given this  approximate response 
kernel, we
investigate the behavior of the slope as a function of $\tau_R$.
Based on our results for the case of two delta functions,
 we expect that chemotaxis be weak if $\tau_R$ is either much smaller than the
delay of the positive peak in the response kernel or much larger than the
delay of the negative peak. We expect optimum chemotaxis for a value of
$\tau_R$ that falls in between the two delays. We verify this prediction in Fig. 
\ref{fig:bilin} (in the linear model). We note that the maximum slope occurs
for a value of $\tau_R$ close to the experimentally recorded 
value of about $1s$.  

\subsection*{Coarse-grained description of chemotaxis as diffusion with drift}

In order to gain insight into our numerical
 results, we developed  a simple coarse-grained  model of chemotaxis.
For the sake of simplicity, we first present the model for a non-adaptive,
singular response 
kernel, $R(t) =\alpha \delta(t-\Delta)$, and, subsequently, we generalize the
model to adaptive response kernels by making use of linear superposition.

\begin{figure}[!ht]
\includegraphics[scale=1.0,angle=0]{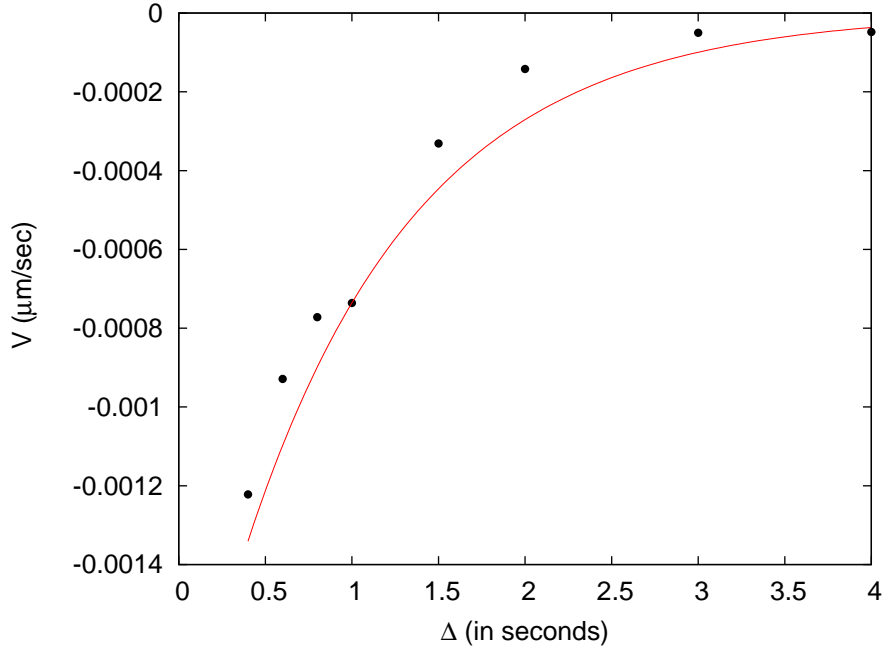}
\caption{ The chemotactic drift velocity, $V$, as a function of $\Delta$, for
the response kernel $R(t) =\alpha \delta (t- \Delta)$. Solid circles:
numerical results. Line: approximate analytical results from \cite{gennes}.  
$\tau_R =1s$  and other numerical parameters as in Fig. \ref{fig:satq5}}
\label{fig:vdel}
\end{figure}

The memory trace embodied by the response kernel induces temporal correlations
in the trajectory of the bacterium. However, if we consider the coarse-grained
motion of the bacterium over a spatial scale that exceeds the typical run
stretch and a temporal scale that exceeds the typical run duration, then we
can assume that it behaves as a Markovian random walker with drift velocity
$V$ and diffusivity $D$. Since the steady-state probability
distribution, $P(x) = P(\Delta, \tau_R,x)$,
  is flat for $\alpha=0$, for small $\alpha$ we can write 
\ber
P &=& P_0 + \alpha {\mathcal P} (\Delta, \tau_R, x) + o(\alpha ^2), \\
D &=& D_0 + \alpha {\mathcal D} (\Delta, \tau_R, x) + o(\alpha ^2), \\
V&=&\alpha {\cal V} (\Delta, \tau_R, x) + o(\alpha ^2).
\eer  
Here, $P_0 = 1/L$ and $D_0= v^2 \tau_R$. 
Since we are neglecting all higher order corrections in
$\alpha$, our analysis is valid only when $\alpha$ is sufficiently small. In
particular, even when $\Delta \gg \tau_R$, we assume that 
the inequality $\Delta /
\tau_R \ll 1/ \alpha$ is still satisfied.
The chemotactic drift velocity, $V$, vanishes if
$\alpha =0$; it is defined as the mean displacement per unit time 
of a bacterium  starting a new run 
at a given location.  Clearly, even in the steady state 
when the current $J$, defined through $\partial_t P =- \partial_x J$, vanishes,
$V$ may be non-vanishing (see Eq. \ref{eq:rw2} below). In general, the non-Markovian
dynamics make $V$  dependent on the initial conditions. However,
in the steady state this dependence is lost and $V$ can be calculated, for
example, by performing a weighted average over the  probability of 
histories of a bacterium. This is the quantity that is of interest
to us. An earlier calculation by de Gennes showed that, if the memory
 preceding the last tumble is ignored, then for a linear
 profile of nutrient concentration the drift velocity is independent of
position and takes the form 
$V= \alpha c v^2 \tau_R \exp(-\Delta / \tau_R)$ \cite{gennes}. While the
calculation applies strictly in a regime with  $\Delta \ll \tau_R$ (because of memory
erasure), in fact its result captures the
behavior well over a wide range of parameters (see Fig. \ref{fig:vdel}).
 To measure $V$
in our simulations, we compute the average displacement of the bacterium
 between two successive tumbles in the steady state,  
 and we extract therefrom the drift velocity.  (For details of the
derivation, see  Supporting Information.) We find that $V$ is negative
for $\alpha <0$ and that its magnitude 
falls off with increasing values of $\Delta$ (Fig. \ref{fig:vdel}). 
We also verify that
$V$ indeed does not show any spatial dependence (data shown in Fig. $S3$ of
Supporting Information). We recall that, in our numerical analysis, we have used
a small value of
 $\alpha$; this results in a low value of $V$. We show below that for an
experimentally measured bilobe
 response kernel, obtained by superposition of singular
response kernels, the magnitude of $V$ becomes larger and
 comparable with experimental values. 

\begin{figure}[!ht]
\includegraphics[scale=1.0,angle=0]{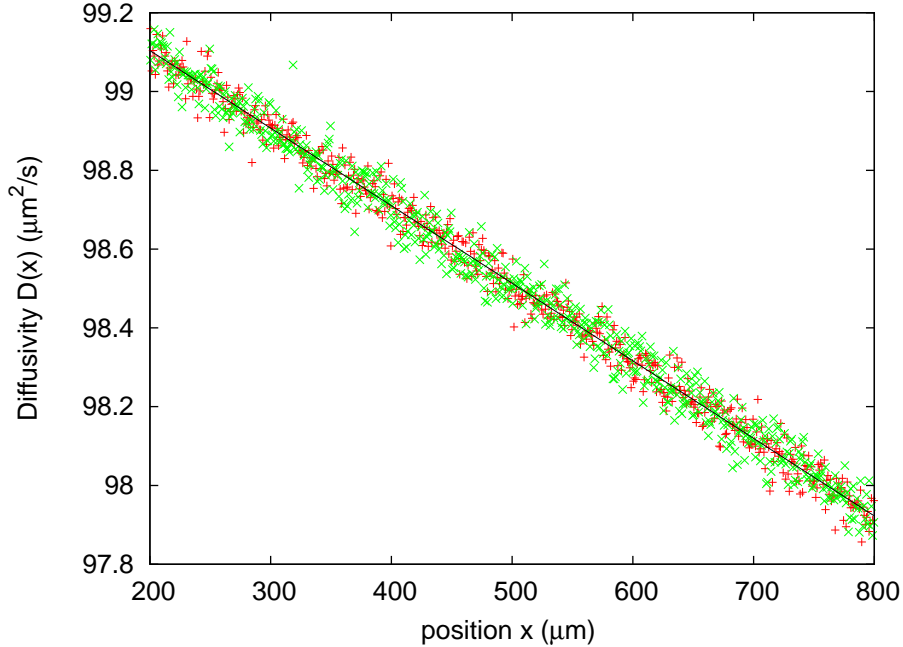}
\caption{ The diffusivity, $D(x)$, as a function of position, $x$, for
 the response
kernel $R(t) =\alpha \delta (t-\Delta)$ with $\Delta = 1s$ ($+$) and $2s$
($\times$).  Instead of plotting $D(x)$ for the entire range of
$x$, we leave out boundary regions to avoid the effect of the reflecting
walls. (From the numerics, the width of the boundary layer is $\sim 80 \mu
m$.) 
 $D(x)$ falls off linearly with $x$ and is independent of $\Delta$. 
 Data fitting yields $D(x)=99.5 - 0.00197 x$ and the coarse-grained model
predicts $D(x)=v^2 \tau_R(1+\alpha c x)$. For the chosen set of parameters,
$v^2 \tau_R =100 \mu m ^2 /s$ and the  $v^2 \tau_R \alpha c = -0.002$. 
 The discrepancy between the numerical and the predicted slopes is due to
higher-order
corrections in $\alpha$, while  discretization of space in simulations
 causes the
slight mismatch in the constant term.  
$\tau_R=1s$ and other numerical parameters are as in Fig. \ref{fig:satq5}. }
\label{fig:dx}
\end{figure}

To obtain the diffusivity, $D$, we first calculate the
effective mean free path in the coarse-grained model. 
 The tumbling frequency of a bacterium is $(1-\alpha c x(t-\Delta))/\tau_R$
 and depends on the details of its past trajectory. In the
coarse-grained model, we replace the quantity $ \alpha c x(t-\Delta)$
by an average $\alpha c \langle x(t-\Delta) \rangle$ over all the trajectories
within the spatial resolution of the coarse-graining. Equivalently, in a
population of non-interacting bacteria, the average is taken over all the
bacteria contained inside a blob, and, hence, $\langle x(t-\Delta) \rangle$
denotes the position of the center of mass of the blob at a time $t-\Delta$ 
in the past. As mentioned above, the drift velocity is
proportional to $\alpha$, so that $\alpha c \langle x(t-\Delta) \rangle
=\alpha c x(t) + O(\alpha^2) $.  
The average tumbling frequency then becomes 
$(1-\alpha c x))/\tau_R$ and, consequently, the mean free path becomes
$\tau_{MFP} = \tau_R /(1-\alpha c x ) \simeq \tau_R (1+\alpha c x)$. 
As a result, the diffusivity is expressed as 
  $ D= v^2 \tau_{MFP} \simeq v^2 \tau_R (1+\alpha c x)$. 
We checked this form against our numerical results (Fig. \ref{fig:dx}).

Having evaluated the drift velocity, $V$, and the diffusivity, $D$, we now
proceed to write down the continuity equation (for a more 
rigorous but less intuitive approach, see \cite{verg}). For a biased random walker on a
lattice, with position-dependent hopping rates  $d^{+}(x)$ and $d^{-}(x)$
towards the right and the left, respectively,
 one has $V= a\left ( d^{+}(x) - d^{-}(x)
\right )$ and $D = a^2  \left ( d^{+}(x) + d^{-}(x) \right )/2 $, where $a$ is
the lattice constant. In the continuum limit,  the temporal evolution of
the probability density is given by a probability current, as  
\be
\partial_t P = -\partial_x J,
\label{eq:rw1}
\ee
 where the current  takes the form 
\be
J =  VP - \partial_x \left ( D P \right ).
\label{eq:rw2}
\ee
 For reflecting boundary condition, $J=0$ in the steady state. This
constraint  yields a steady-state slope
\be
\beta = \alpha \partial_x {\cal P}= \alpha \frac{P_0}{D_0} \left ( 
{\cal V} -\partial_x {\cal D} \right ) = 
\frac{\alpha {\cal V} }{L v^2 \tau_R} -\frac{\alpha c }{L}
\label{eq:slope}
\ee
for small $\alpha$. We use our measured values for $V$ and $D$  (Figs. 
\ref{fig:vdel} and
\ref{fig:dx}), and compute the slope using Eq. \ref{eq:slope}.
(For details of the measurement of $V$, see Supporting Information.)
 We compare our analytical and numerical results in Fig.
\ref{fig:satq5}, which exhibits close agreement.

According to Eq. \ref{eq:slope},  steady-state chemotaxis results from a 
competition between drift motion and diffusion. For $\alpha < 0$, the drift
motion is directed toward regions with a lower
 nutrient concentration and hence opposes chemotaxis. Diffusion is spatially
dependent and becomes small
  for large nutrient concentrations (again for $\alpha <0$), thus
 increasing the effective residence
time of the bacteria in favorable regions. For large values of 
$\Delta$, the drift velocity vanishes and
one has a strong chemotaxis as
$\Delta$ increases (Fig. \ref{fig:satq5}).
Finally, for $\Delta =0$, the calculation by de Gennes 
yields $V=\alpha c v^2 \tau_R$ which
exactly cancels the spatial gradient of $D$  (to linear order in
$\alpha$), and there is no accumulation \cite{gennes,lars}.

 These conclusions are easily generalized to adaptive response functions. 
 For $R(t)=\alpha \delta (t-\Delta_1) - \alpha \delta (t-\Delta_2)$, within
the linear response regime, the effective
drift velocity and diffusivity can be constructed by simple linear 
superposition: The drift velocity reads $V = \alpha
{\cal V} (\Delta_1) - \alpha {\cal V} (\Delta_2) $. Interestingly, the
 spatial dependence of $D$ cancels out 
and $D = D_0 = v^2 \tau_R$. The resulting slope then depends on the
drift only and is calculated as 
\be
\beta =  \frac{\alpha}{L v^2 \tau_R} \left (  {\cal V}(\Delta_1) - 
{\cal V}(\Delta_2) \right ).
\label{eq:2del}
\ee  
In this case, the coarse-grained model is a simple biased random walker with
constant diffusivity. For $\Delta_1 < \Delta_2$ and $\alpha >0$, the net
velocity, proportional to 
 $\alpha \left (  {\cal V}(\Delta_1) - {\cal V}(\Delta_2) \right )$,
 is positive and gives rise to a  favorable chemotactic response, according to
 which bacteria
accumulate in regions with high food concentration. Moreover, 
the slope increases as the separation between $\Delta_1$ and $\Delta_2$ grows. 
 We emphasize 
 that there is no incompatibility between strong steady-state
chemotaxis and large drift velocity. In fact, in the case of an
adaptive response function, strong chemotaxis occurs only when the
drift velocity is large.

For a bilobe response kernel, approximated by a superposition of many
delta functions (Fig. \ref{fig:r}), the slope, $\beta$,
 can be calculated similarly 
and in Fig. \ref{fig:bilin} we compare our calculation to 
the simulation results. We find close agreement in the case of a linear model
with a bilobe response kernel and, in fact, also in the case of a non-linear
model (see Supporting Information).

 The experimental bilobe response kernel $R(t)$ is a smooth
function, rather than a finite sum of singular kernels over a set of discrete 
 $\Delta$ values (as in Fig. \ref{fig:r}). Formally, we
integrate singular kernels over a continuous range of $\Delta$ to obtain a 
smooth response kernel.  If we then integrate 
the expression for the drift velocity obtained by de Gennes, according to this
 procedure, we find an overall drift velocity  $V \sim 0.3 \mu m/s $, for the
concentration gradient considered  ($\nabla c=0.001 \mu m^{-1}$). By scaling up
 the concentration gradient
by a factor of $\kappa$, the value of $V$ can also be scaled up by $\kappa$
and can easily account for the experimentally measured velocity range.

\section*{Discussion}

We carried out a detailed analysis of steady-state bacterial
chemotaxis in one dimension. The chemotactic performance in the case of a 
linear concentration profile  of the chemoattractant, $c(x)=cx$, 
was measured as 
 the slope of the bacterium probability density profile in the steady state. 
For a singular 
impulse response kernel, $R(t) = \alpha \delta (t-\Delta)$, the slope was a  
 scaling function of $\Delta / \tau_R$, which vanished at the
origin, increased monotonically, and saturated at large argument. To
understand these results we proposed a simple coarse-grained model  in
which bacterial 
motion was described as a biased random walk with drift
velocity, $V$, and diffusivity, $D$. We found that for small
enough values of $\alpha$, $D$ was independent of $\Delta$
and varied linearly with nutrient concentration. By contrast, 
$V$ was spatially uniform and its value decreased
monotonically with $\Delta$ and vanished for $\Delta \gg \tau_R$. We presented
a  simple
formula for the steady-state
slope in terms of $V$ and $D$. The prediction of our coarse-grained model 
 agreed closely with our numerical results. Our description
is valid when
$\alpha$ is small enough, and all our results are derived to linear order in
$\alpha$. We assume $\Delta / \tau_R \ll 1/\alpha$ is always satisfied.

Our results for an impulse response kernel can be easily generalized 
to the case of response kernels with 
arbitrary shapes in the linear model. For an adaptive
response kernel, the spatial dependence of the diffusivity, $D$,
cancels out but a positive drift velocity, $V$, ensures bacterial
accumulation in regions with high nutrient concentration, in the steady state.
In this case, the slope is
directly proportional to the drift velocity. As the delay 
between the positive and negative peaks of the response kernel grows,
the velocity increases, with consequent stronger chemotaxis.

Earlier studies of chemotaxis \cite{schnitzer,coarse,mittal,blythe}
put forth a coarse-grained model different from ours. In the
model first proposed by Schnitzer  for a single chemotactic
 bacterium \cite{schnitzer}, he argued that, in order to  obtain favorable
bacterial accumulation, 
tumbling rate and ballistic speed 
of a bacterium must  both depend on the direction of its motion. In his case, 
the continuity equation reads
\be
\partial_t P = \partial_x \left [ \frac{\gamma_L v_R -\gamma_R
v_L}{\gamma_L+\gamma_R} P - 2\frac{v_R+v_L}{\gamma_R+\gamma_L} \partial_x
\left ( \frac{v_R v_L}{v_R+v_L} P \right) \right ],
\label{eq:sch}
\ee
where $v_{L(R)}$ is the  ballistic speed and $\gamma_{L(R)}$ is
the tumbling frequency 
of a bacterium moving toward the left (right). For {\sl E. coli}, as discussed
above, $v_L= v_R =v$, a constant independent of the location. 
In that case, Eq. \ref{eq:sch}  predicts that in order to
have a chemotactic response in the steady state,
 one must have a non-vanishing drift velocity, {\sl i.e.}, $ \left ( \gamma_L v_R -
\gamma_R v_L \right )/ (\gamma_L + \gamma_R ) \neq 0$. This 
contradicts our findings for non-adaptive response kernels, according to which
a drift velocity only hinders the  chemotactic response. The
spatial variation of the diffusivity, instead, causes the chemotactic
accumulation. This is not captured by Eq. \ref{eq:sch}.
In the case of adaptive response kernels, the diffusivity 
becomes uniform while the drift velocity is positive, 
 favoring chemotaxis. Comparing the expression of the flux, $J$, obtained
from Eqs. \ref{eq:rw1} and \ref{eq:rw2} with that from Eq. \ref{eq:sch}, 
and matching  the respective coefficients of $P$ and
$\partial_x P$, we find $D=2v_R v_L/(\gamma_R + \gamma_L) $ and 
$V = (\gamma_L v_R -\gamma_R v_L)/(\gamma_L + \gamma_R)$. As we argued above
in  discussing the coarse-grained model for adaptive
response kernels, both $D$ and $V$ are spatially independent. This puts 
strict restrictions on the spatial dependence of $v_{L(R)}$ and
$\gamma_{L(R)}$. For example, as in {\sl E. coli} chemotaxis  
$v_L= v_R=v$, our coarse-grained description is recovered only 
if $\gamma_L$ and $\gamma_R$ are also independent of $x$.

We  comment on a possible origin of the discrepancy between our work and
earlier treatments. In Ref. \cite{schnitzer}, a
continuity equation was derived for the coarse-grained probability density
of a bacterium, starting from a pair of 
approximate master equations for the probability density of a right-mover and 
a left-mover,  respectively.
As the original process is non-Markovian, one can expect a master
equation approach to be valid only at scales that exceed the scale over which
spatiotemporal correlations in the behavior of the bacterium are significant.
In particular, a biased diffusion model can be viewed as legitimate only if
the (coarse-grained) temporal resolution allows for multiple runs and tumbles.
If so, at the resolution of the coarse-grained model, left- and
right-movers become entangled, and it is not possible to perform a
coarse-graining procedure on the two species separately. Thus one cannot
define probability densities for a left- and a right-mover that evolves in a
Markovian fashion. In our case, left- and right-movers are coarse-grained
simultaneously, and the total probability density is Markovian. Thus, our
diffusion model differs from that of Ref. \cite{schnitzer} because it
results from a different coarse-graining procedure. The model proposed in Ref. 
\cite{schnitzer} has been used extensively to investigate
 collective behaviors of 
{\sl E. coli} bacteria such as pattern formation \cite{coarse,mittal,blythe}. 
It would be worth asking whether the new coarse-grained
description  can shed new light on bacterial
 collective behavior.

\section*{Acknowledgments}
It is a pleasure to thank Damon Clark and Massimo Vergassola for their comments 
on the manuscript.

\newpage

\section*{Supporting Information}

\subsubsection*{Small-argument behavior of $F\left ( \Delta /\tau_R \right )$}

Here, we argue that for small $\Delta / \tau_R$ the function $F\left ( \Delta
/\tau_R \right )$ is linear. First, note that 
for an impulse response kernel, $R(t)=\alpha \delta(t-\Delta)$ the tumbling
probability during the interval
 $[t,t+dt]$ is $ \left [ 1-\alpha c x(t-\Delta)
\right  ] dt/\tau_R$. For small $\Delta$, we can assume that no tumbling
occurs 
during the interval $[t-\Delta,t]$. Then the effective tumbling rates 
 become $\left [ 1-\alpha c (x-v\Delta) \right ]/\tau_R$  for
right-movers and $\left [ 1-\alpha c (x+v\Delta) \right ]/\tau_R$  for
left-movers. Based on this observation,
we can write a pair of master equations that govern the densities
 of left-movers
and right-movers, and we can solve them in the steady state. 
 We obtain the slope, $\beta$, expressed as 
\be
\beta =  -\frac{2 \alpha c \Delta }{L \tau_R }. 
\ee
Thus, in the limit $\Delta \ll \tau_R$ the scaling function 
$F(\Delta /\tau_R)$ depends linearly on its argument.  

\subsubsection*{ Chemotactic drift velocity as a function of position }
\begin{figure}
\includegraphics[scale=1.0,angle=0]{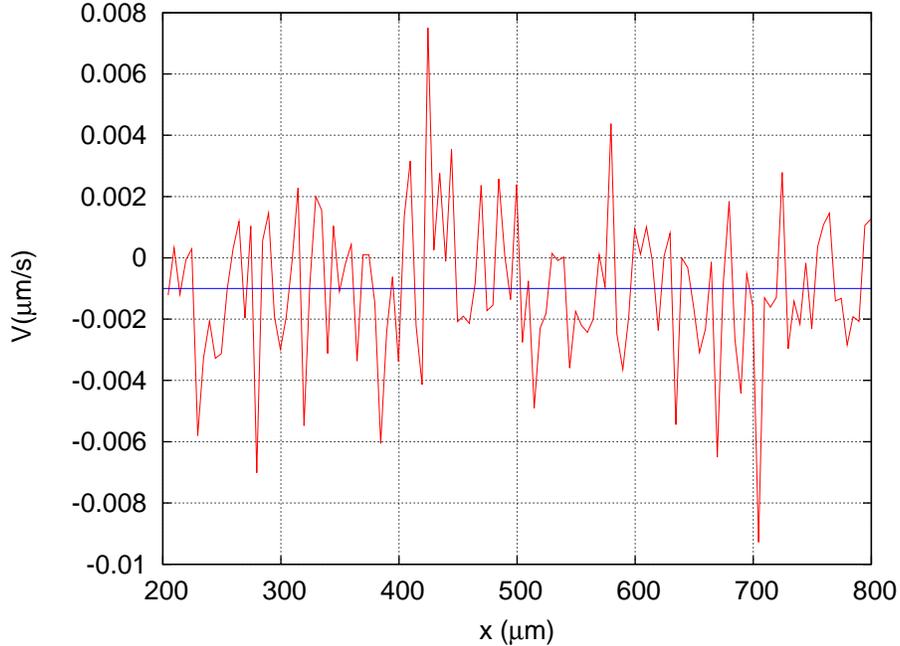}
\caption{\it Drift velocity, $V$, as a function of position, $x$,  
for the case of a singular 
response kernel $R(t) =\alpha \delta (t-\Delta)$ (red line). 
 Instead of showing the plot for the entire range of
$x$, we leave out boundary regions to discard the effect of the reflecting 
walls. Our numerics show that the width of the boundary layer is 
$\sim 80 \mu m$. Here, 
$L=1000 \mu m$, $q=1.0$, $\tau_R= 1s$, $\Delta = 1s$, $\tau_T=0$, 
$v=10 \mu m /s$, $\alpha = -0.1$, $c=0.001 {\mu m}^{-1}$. Based on the data 
shown, the drift velocity is $0.001 \pm 0.0001 \mu m /s $ (blue line). }
\label{fig:vx}
\end{figure}

Here, we show
  a sample of our numerical results on the drift velocity $V$
 as a function of
position. We measure the average displacement of a
bacterium in the steady state (see the corresponding 
discussion in the paper),
 and we compute $V$ therefrom. Specifically, 
we perform the following procedure. We denote by $m_L(x)$ and
$m_R(x)$ the total number of leftward and rightward runs, respectively, 
 that initiate at the position $x$, at a time $t$, for a population of
non-interacting bacteria.
These quantities are well-defined in the context of our numerics because 
space is discretized.
  Furthermore, let $S_L(x)$ and  $S_R(x)$ be the total
leftward and rightward displacement, respectively, of the bacteria that
undergo  these runs. The
average displacement  per run is then given by
  $[S_R(x) - S_L(x)]/[m_R(x) + m_L(x)]$.  Of course, the choice of the 
time scale as the duration
 of a run is arbitrary and other choices are equally valid. 
To leading  order, 
this average displacement is linear in $\alpha$. 
The drift velocity $V$, to order
$\alpha$, is then obtained by dividing this average displacement by $\tau_R$. 
(Note that any $O(\alpha)$ correction to $\tau_R$ leads to  $O(\alpha ^2)$ 
correction to $V$, which we ignore here.) We find that up to the
noise present in our numerical measurement,
 $V$ does not  display any dependence on  
position. An example is illustrated in Fig. \ref{fig:vx}. 

\subsubsection*{Dependence of $V$ on the turning probability $q$ }

\begin{figure}
\includegraphics[scale=1.0,angle=0]{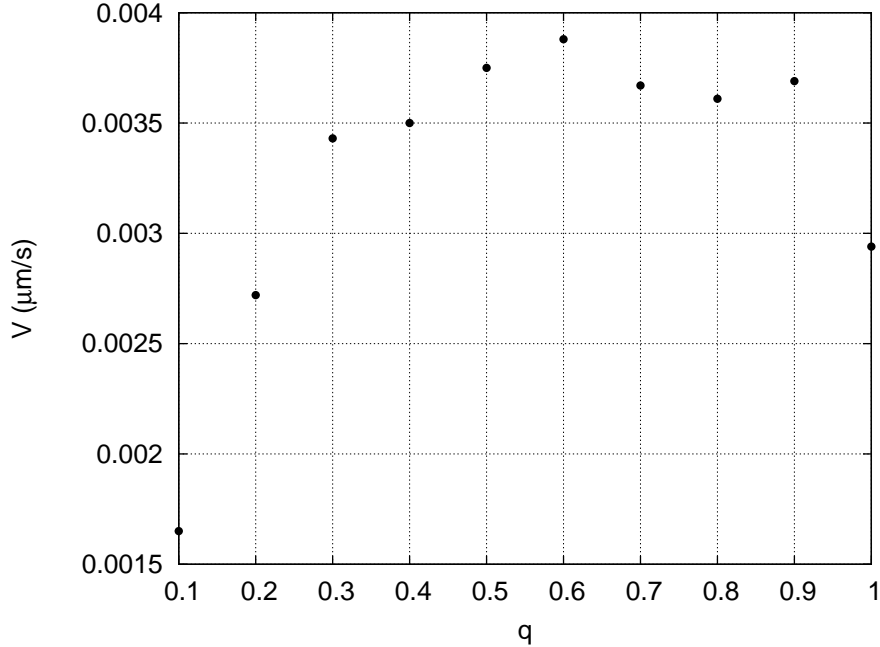}
\caption{\it The drift velocity, $V$, as a function of the turning probability,
$q$, for the case of a singular adaptive 
response kernel $R(t) =\alpha \delta (t-\Delta_1) - \alpha \delta
(t-\Delta_2)$. Here, $L=1000 \mu m$, $\tau_R= 1s$, $\Delta_1 = 0.5s$, 
$\Delta_2 =1.5s$, $\tau_T=0$, 
$v=10 \mu m /s$, $\alpha = 0.1$, $c=0.001 {\mu m}^{-1}$. }
\label{fig:vq}
\end{figure}

In our model, $q$ denotes the turning probability, {\sl i.e.}, the probability
that the run direction inverses after a tumble.
 Our numerical explorations indicate that changing the value of $q$
does not affect the qualitative behavior of the system. 
However, the numerical value of the drift 
velocity, and the value of $\beta$  depend on $q$. In
Fig. \ref{fig:vq}, we exhibit the variation of $V$ as a function of
 $q$, in the case of an adaptive response function.

\subsubsection*{Results in the non-linear model}

Some earlier experiments indicate that bacteria modulate their run 
durations in response to a positive concentration gradient, but 
not to a negative one. In order to incorporate this feature in our
model, we have to go beyond the linear response regime. In the non-linear
model, whenever the linear functional (Eq. 2) becomes negative, it is replaced 
by $0$. This is the only difference with the linear
model. Thus, for a purely positive response kernel the non-linear model
behaves identically to the linear model, while for a purely negative response
kernel the non-linear model displays no chemotaxis whatsoever. Hereafter, we
examine only adaptive response kernels with balanced positive and negative
contributions.

\begin{figure}
\includegraphics[scale=1.0,angle=0]{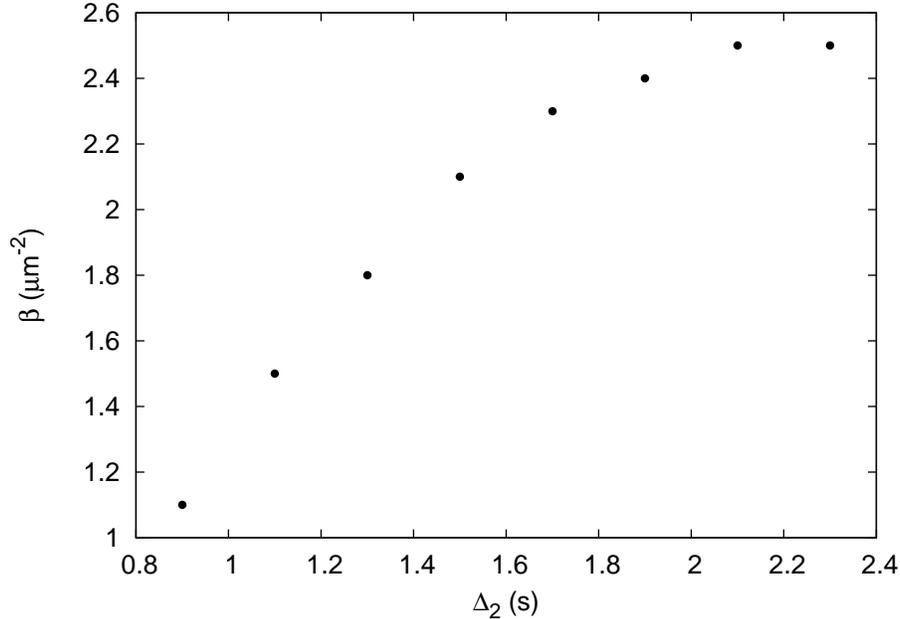}
\caption{The slope, $\beta$ (scaled by a factor of $10^8$),
 as a function of $\Delta_2$, for $\Delta_1 = 0.5
s$, in a non-linear model with  balanced  response kernel, $R(t)=\alpha 
\delta (t-\Delta_1) - \alpha \delta(t-\Delta_2)$. As in the linear
model, $\beta$ increases with the difference of $\Delta_1$ and
$\Delta_2$.
Here, $\tau_R=1$ sec, $q=0.4$, $\alpha = 0.1$, $L=1000 \mu m$, $c=0.001 \mu
m^{-1}$, $v=10 \mu m/s$. }
\label{fig:nlin1}
\end{figure}

\begin{figure}
\includegraphics[scale=1.0,angle=0]{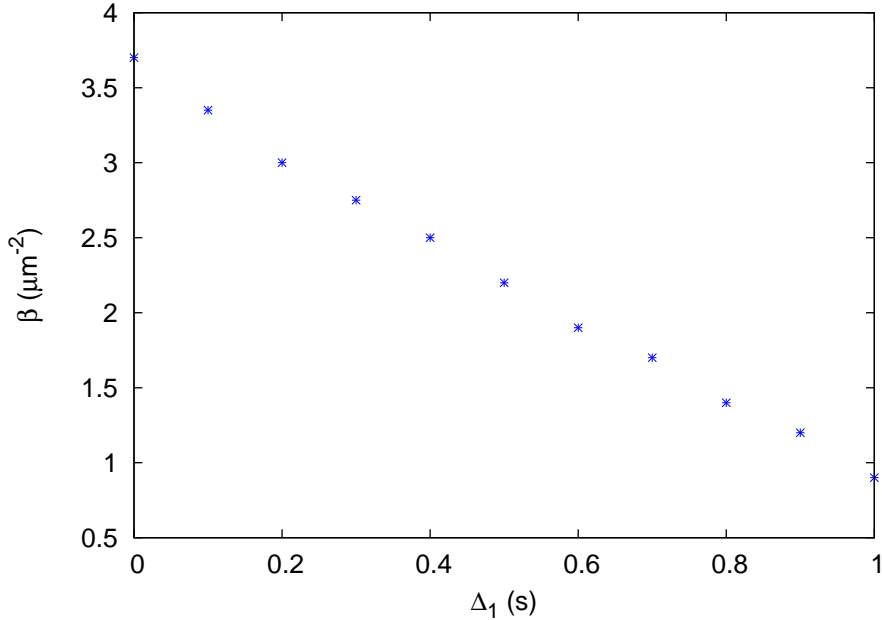}
\caption{ The slope, $\beta$ (scaled by a factor of $10^8$),
 as a function of $\Delta_1$, for $\Delta_2 = 1.5
s$, in a non-linear model with balanced response kernel, $R(t)=\alpha 
\delta (t-\Delta_1) - \alpha \delta(t-\Delta_2) $. Numerical
 parameters are as in Fig. \ref{fig:nlin1}. }
\label{fig:nlin2}
\end{figure}

We first consider the idealized response kernel made of the superposition of 
positive and  negative delta functions, $R(t) = \alpha \delta (t-\Delta_1)
-\alpha \delta (t-\Delta_2)$. Simulations show that
many qualitative features of the linear model
still hold in the non-linear model.
 The scaling form valid in the linear case breaks down in the non-linear case, but 
the slope $\beta$ increases with the separation between $\Delta_1$ and $\Delta_2$
 and ultimately saturates  to a non-vanishing value.
 Figures \ref{fig:nlin1} and
 \ref{fig:nlin2} display results of simulations. As expected, strong
chemotaxis occurs when $\Delta_1=0$ and $\Delta_2$ is substantially
larger than $\tau_R$. We have also verified that tumbling does not have much
of an effect on the slope.

\begin{figure}
\includegraphics[scale=1.0,angle=0]{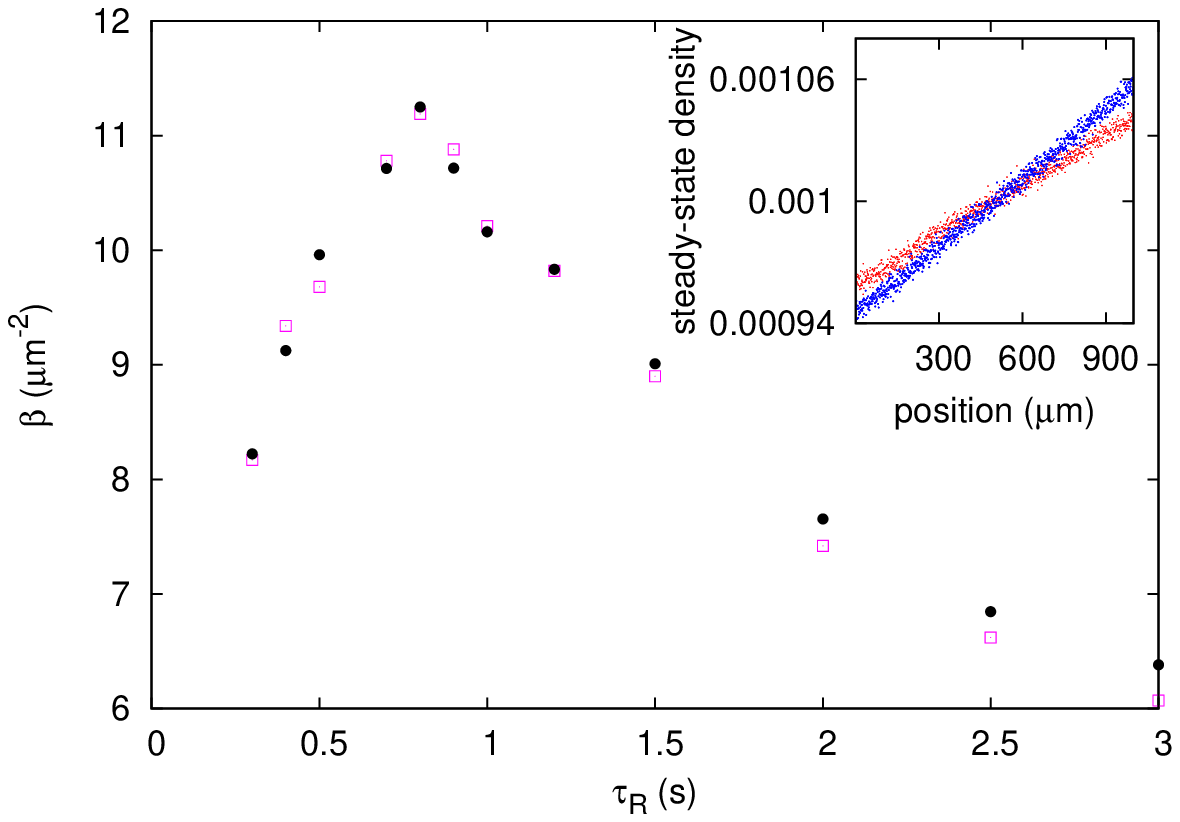}
\caption{ The slope, $\beta$ (scaled by a factor of $10^8$),
 as a function of $\tau_R$, in a non-linear
model with the experimental bilobe response kernel of Fig. $1$ in the paper.
 Open
squares: numerical results from simulations. Solid circles: prediction of
the coarse-grained model. Numerical parameters as in Fig. $3$ in the paper. The
inset shows the steady-state density profiles of the bacterial population for
$\tau_R=0.3,0.8 s$ (red and blue curve), respectively. }
\label{fig:binlin}
\end{figure}

\begin{figure}
\includegraphics[scale=1.0,angle=0]{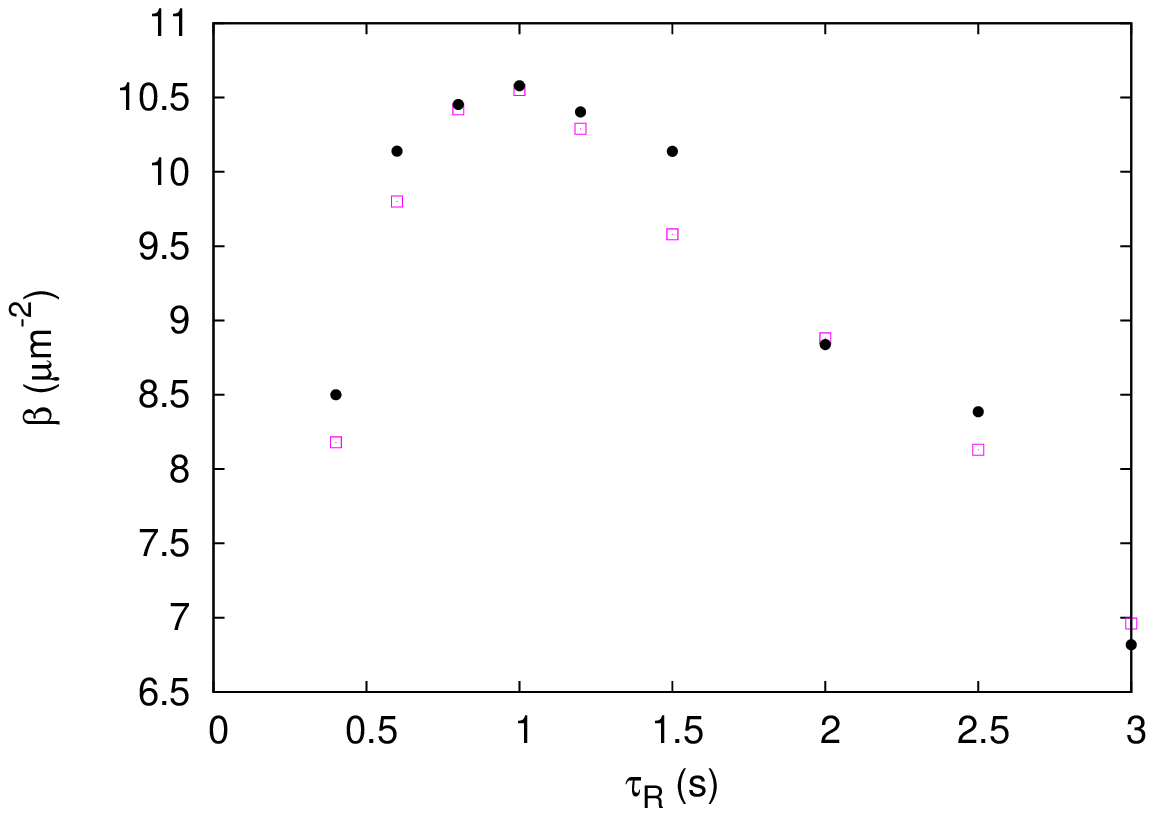}
\caption{The slope, $\beta$ (scaled by a factor of $10^8$),
 as a function of $\tau_R$, in a non-linear
model with the experimental bilobe response kernel of Fig. $1$ in the paper. Here, $q=0.5$ and the other numerical parameters are as in Fig. $3$ 
in the paper. Comparison with Fig. \ref{fig:binlin} shows that the position of
the maximum depends on the value of $q$. 
Open squares: numerical results from simulations.
 Solid circles: prediction of the coarse-grained model.}
\label{fig:q5nlin}
\end{figure}

In the experimental case of a bilobe response kernel (Fig. 1 in the paper),
 we find that strong chemotaxis occurs when $\tau_R$ lies
between the positive and the negative peaks of the response kernel, as 
found in the linear case.
 For smaller or larger values of $\tau_R$, chemotaxis becomes weak. Figures 
\ref{fig:binlin} and  \ref{fig:q5nlin} show our numerical results for $q=0.4$
and $q=0.5$, respectively. We   
note that, in both plots
  the value of $\tau_R$ for which the slope is maximum falls
close to the experimental value of about $1s$. However, the exact position of
the maximum depends on $q$. For $q=0.4$ (Fig.\ref{fig:binlin}) maximum 
occurs at $\tau_R \simeq 0.8s$, while for $q=0.5$ (Fig. \ref{fig:q5nlin}) the
maximum occurs at $\tau_R
\simeq 1s$.  

\subsubsection*{ Chemotaxis with non-vanishing tumbling durations}

During a tumbling event the bacterium rotates about itself in a 
random fashion without any significant displacement. In a homogeneous
nutrient concentration the average tumbling duration is $0.1 s$, which is much
smaller than the average run duration of $1 s$. In the
 steady state one therefore
expects that the bacterium spends only  a fraction
 $\tau_T/\tau_R \simeq 0.1$ of the 
 time in the tumbling state.  For this reason, studies of
chemotaxis often assume instantaneous tumbling.

It was shown recently that the existence of non-vanishing 
 tumbling duration can
yield interesting results: even with a punctual
 response kernel, $R(t) = \alpha \delta (t-\Delta)$ with $\Delta =0$, 
{\sl i.e.}  a memoryless bacterium, one can observe  a
chemotactic response (Kafri {\sl et al.}, 2008). Here, we  provide a
simplified derivation of the steady-state density profile in this Markovian
limit. The result will prove useful also for the analysis of the non-Markovian
case with $\Delta \neq 0$.

\begin{figure}
\includegraphics[scale=1.0,angle=0]{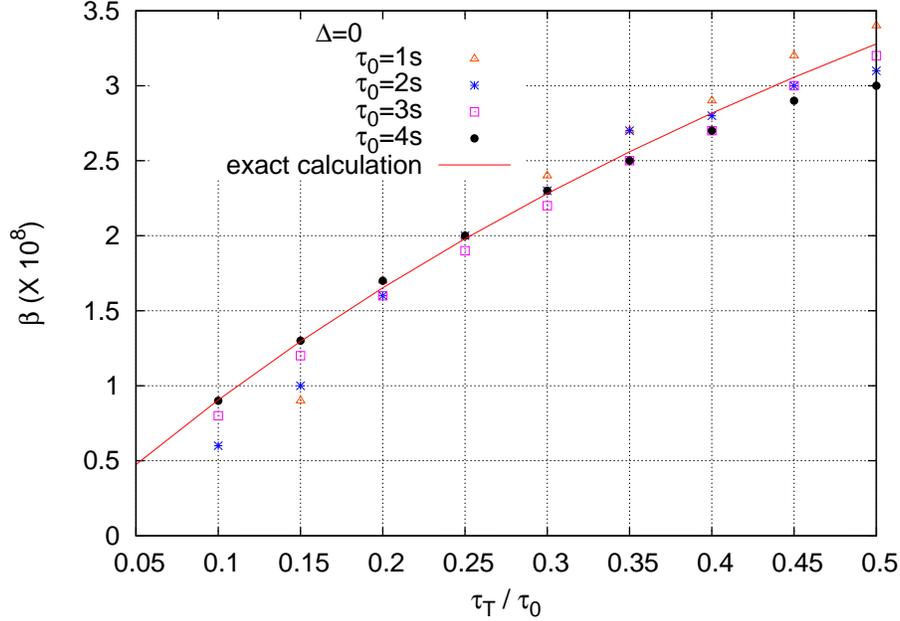}
\caption{\it The steady-state slope $\beta$ as a function of $\tau_T/\tau_0$ 
for $\Delta =0$ and unmodulated tumbling. The solid line corresponds to the exact result from Eq. \ref{eq:notum}. Here we have used $L=1000 \mu m$,
 $\alpha =-0.1$, $q=0.4$, $c=0.001 \mu m^{-1}$, $v=10 \mu m /sec$.} 
\label{fig:notum}
\end{figure}

Let $L(x,t)$ and
$R(x,t)$ be the density of left-movers and right-movers, respectively, at
 location $x$ and time $t$. We denote by
 $T^R (x,t)$ and $T^L (x,t)$ the densities of
tumblers that were moving to the right and left, respectively, before
tumbling. For $\Delta =0$, the time evolution of these quantities can be 
described by master equations. In the case in which tumble durations are not
modulated and tumble-to-run switches always occur at a fixed rate
$1/\tau_T$, the master equations read
\ber
\nonumber
\frac{\partial R (x,t)}{\partial t} =& -v\partial_x R (x,t) + T^R (x,t)
\frac{(1-q)}{\tau_T} \\
& +  T^L (x,t) \frac{q}{\tau_T} - R(x,t) \frac{1-\alpha c
x}{\tau_0},& \\
\nonumber
\frac{\partial L (x,t)}{\partial t} =& v\partial_x L (x,t) + T^L (x,t)
\frac{(1-q)}{\tau_T} \\
& + T^R (x,t) \frac{q}{\tau_T} - L(x,t) \frac{1-\alpha c
x}{\tau_0}, \\
\frac{\partial T^R (x,t)}{\partial t} =& R(x,t) \frac{1-\alpha c x}{\tau_0} -
T^R(x,t)\frac{1}{\tau_T},\\
\frac{\partial T^L (x,t)}{\partial t} =& L(x,t) \frac{1-\alpha c x}{\tau_0} -
T^L(x,t)\frac{1}{\tau_T}. 
\eer
We consider perfectly reflecting boundary conditions at $x=0$ and $x=L$. This
implies that, in the steady state, we must have $R(x)=L(x)$. The steady-state 
(total) density at location $x$ then becomes 
\ber
\nonumber
N(x)=&R(x)+L(x)+T^R(x)+T^L(x) \\
=& \frac{2}{L}\frac{1}{2+\frac{\tau_T}{\tau_0}
(2-\alpha)} \left (1+\frac{\tau_T}{\tau_0} (1-\alpha c x) \right ).
\eer
Therefore, the slope of the steady-state density profile is given by 
\be 
\beta = G \left (  \frac{\tau_T}{\tau_0} \right ) = -2\alpha c \frac{1}{L}
\frac{\tau_T}{\tau_0}  \frac{1}{2+\frac{\tau_T}{\tau_0}(2-\alpha)}.
\label{eq:notum}
\ee
In Fig. \ref{fig:notum}, we compare this result with the slope measured in
simulations. This plot demonstrates that, even in the 
absence of any memory or modulation of tumbling durations, 
it is possible to obtain chemotaxis in the steady state.  

\begin{figure}
\includegraphics[scale=1.0,angle=0]{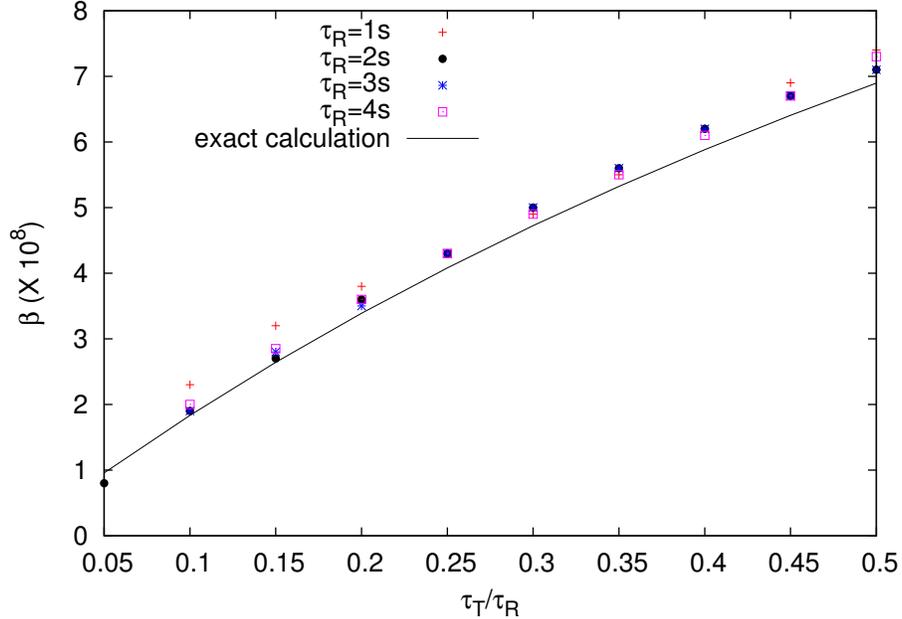}
\caption{\it The slope, $\beta$, as a function of 
 $\tau_T/\tau_R$ for $\Delta =0$
and modulated tumbling.  
 Simulation parameters are $L=1000 \mu m$, $\alpha = -0.1$,
$q=0.4$, $c=0.001 \mu m^{-1}$, $v=10 \mu m /s$. }
\label{fig:tum}
\end{figure}

This result for the slope is slightly different for the case in which
 tumbling durations are
modulated: then, $\tau_T$
in the above master equations is replaced by $\tau_T/(1+\alpha c x)$. 
Solving for the steady state, we find a total density 
\ber
\nonumber
N(x)=&R(x)+L(x)+T^R(x)+T^L(x) \\
=&\frac{1}{L\left ( 1+\frac{\tau_T}{\tau_0} \right
)} \left ( 1+\frac{\tau_T}{\tau_0}\frac{1-\alpha c x}{1+\alpha c x} \right ).
\label{eq:nx}
\eer
For $\alpha c x \ll 1$, this is approximated by a linear form and the
slope becomes 
\be 
\beta = G \left ( \frac{\tau_T}{\tau_0} \right ) = -\frac{ 2 \alpha c}{L}
\frac{\tau_T}{\tau_0} \frac{1}{1+\frac{\tau_T}{\tau_0}(1-\alpha)}.
\label{eq:tum}
\ee
We compare this analytical result with simulations in Fig. \ref{fig:tum}, which
shows a systematic deviation for large argument. We have verified that 
 this mismatch originates in 
 the linearizing approximation step from Eq. \ref{eq:nx} to Eq. \ref{eq:tum}.

\begin{figure}
\includegraphics[scale=1.0,angle=0]{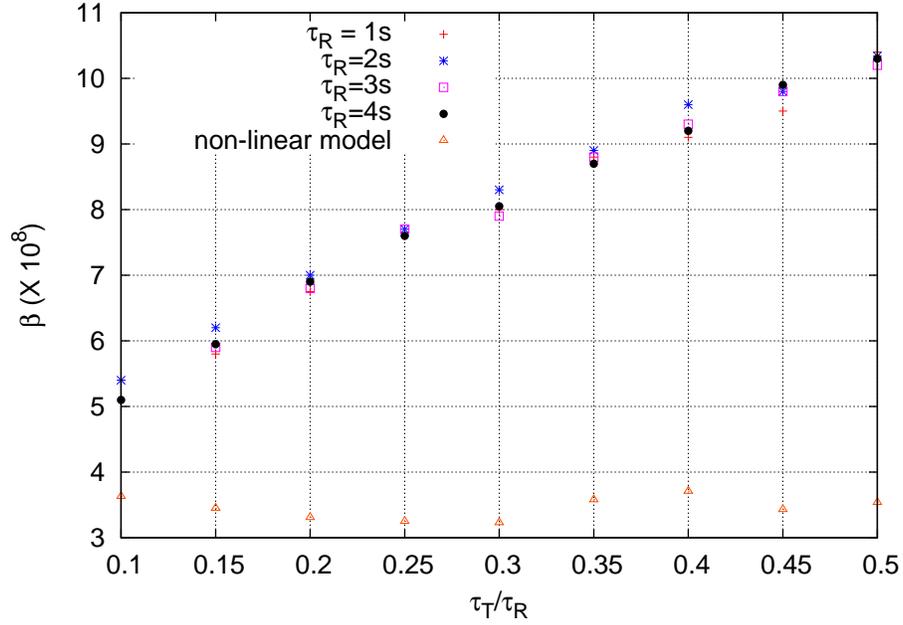}
\caption{\it The scaling collapse of the slope, $\beta$, as a function of 
$\tau_T / \tau_R$ for 
fixed value of $\Delta/\tau_R $. We have used $\Delta/\tau_R= 0.5$ here.
 The other simulation parameters are as in Fig. \ref{fig:tum}. We also
plot the slope, $\beta$, in the non-linear model with 
$R(t)=\alpha \left [ \delta (t-\Delta_1) - \delta (t-\Delta_2) \right ]$,
$\Delta_1=0s$ and $\Delta_2=1.5s$ and $\tau_R=1s$: $\beta$
does not show any significant dependence on $\tau_T$.}
\label{fig:delscale}
\end{figure}

For $\Delta \neq 0$ and when both runs and tumbles are modulated,
we measure the density profile in numerical simulations. For
$R(t)=\alpha \delta (t-\Delta)$, our numerics indicate that
 the steady-state slope, $\beta$, is
a sum of the Markovian component, $G$, defined in  Eq. \ref{eq:tum},
 and a non-Markovian component, $F$, which depends on $\Delta / \tau_R$ but is
independent of $\tau_T$: 
\be
\beta = F \left ( \frac{\Delta}{\tau_R} \right ) + G \left (
\frac{\tau_T}{\tau_R} \right ). 
\label{eq:beta}
\ee
In Fig. \ref{fig:delscale}, we exhibit this scaling form as a
function of $\tau_T / \tau_R$ for a fixed, non-vanishing
 value of  $\Delta /\tau_R$.
Our results suggests that $\beta$ is made up of
 two contributions: one from the modulating
runs, encoded in $F$, and one from  non-instantaneous
 tumbles, encoded in $G$. The latter contribution is
independent of $\Delta$. 

Finally, we can infer more general results from the simple form 
 in Eq. \ref{eq:beta}. In particular, for
an  adaptive response function in the linear model, 
the positive and negative parts of
the response function cancel out the effect of  non-vanishing tumble
durations.
 In this case, the steady-state slope,
 $\beta$, becomes independent of $\tau_T$. Interestingly, we find the
same  applies even in the non-linear model (see Fig. \ref{fig:delscale}).


\begin{thebibliography}{99}
\bibitem{footnote} The bacteria can also be repelled by
chemorepellants and tend to accumulate in low chemical concentration. It
is straightforward to reinterprete all our results and discussions for this 
case.
\end{thebibliography}
\end{document}